\documentclass[aps,prb,twocolumn,groupedaddress,floatfix,showpacs]{revtex4-2}

\usepackage{graphicx}
\usepackage{amsmath}
\usepackage{subcaption}
\usepackage{color}

\definecolor{green}{rgb}{0,0.5,0}

\begin{document}


\title{Plasmon-enhanced quadrupole transitions of Rydberg excitons in Cu$_2$O.}


\author{David Ziemkiewicz}
\email{david.ziemkiewicz@utp.edu.pl}
\author{Sylwia Zieli\'{n}ska-Raczy\'{n}ska}
 \affiliation{Department of
 Physics, Technical University of Bydgoszcz,
\\ Al. Prof. S. Kaliskiego 7, 85-789 Bydgoszcz, Poland}


\date{\today}




\begin{abstract}
A mechanism for amplification of weak quadrupole transitions of Rydberg excitons in Cu$_2$O by a plasmonic nanostructure is investigated. The theoretical description of exciton-plasmon interaction is presented and a field averaging approximation is proposed for calculations involving large excitons. Various types of copper, silver and gold-based plasmonic nanostructures are considered and their potential for enhancing quadrupole transitions is evaluated. 
\end{abstract}



\maketitle

\section{Introduction}
An exciton is a bound electron-hole state in a semiconductor that shares many properties with hydrogen atom. The valence band hole and conduction band electron are bound by Coulomb interaction, forming a neutral quasiparticle. Rydberg excitons are characterized by particularly large principal quantum number $n$, just like Rydberg atoms. Since the observation of states with $n>20$ in Cu$_2$O \cite{Kazimierczuk}, they have become a subject of intensive research due to their unique properties originating from the large value of $n$ \cite{Heck2017}. One of their outstanding features is that their radius size  $r$ is  proportional to $n^2$, which makes their interactions with spatially varying fields and nanostructures particularly interesting. The complicated, nonlocal\cite{Takahata2018} and often highly nonlinear\cite{Morin2022} behavior of excitons makes them one of the most promising systems for compact, solid-state quantum computing applications. To this end, the technology of fabricating Cu$_2$O nanostructures is developing rapidly \cite{Hamid,Steinhauer,Takahata2018}.

Plasmonic nanostructures can be used to enhance two-photon absorption processes or weak quadrupole emission due to their ability to concentrate electromagnetic fields at nanoscale \cite{Karolina}. 
The enhancement results from localized surface plasmon resonances, which appear when the electrons in the metal oscillate collectively in response of an external light source.

Among many types of nanostructures, plasmonic nanoantennas recently attracted attention in the context of increasing the probability of selected quadrupole transitions involving exciton energy levels \cite{Neubauer}. Coupled plasmon-excitons, called plexcitons \cite{Karademir14}, have been studied in various semiconductors \cite{Lee16,Cao18,Goncalves18}, but the interaction of plasmons with large Rydberg excitons is a relatively novel field which may lead to a plethora of applications. For example, higher harmonic generation has been demonstrated with Rydberg atoms in plasmon field \cite{Tikman2016}; excitons in Cu$_2$O could provide a solid-state analogue for such a system.

The general concept of enhancement of quadrupole transitions with plasmons is based on the fact that plasmonic structures are capable of focusing light into very small spots. Such a highly focused field is characterized by a particularly large electric field gradient. This, in turn, affects quadrupole and higher order multipole transitions which are usually exceptionally weak. For example, the use of highly focused field in a dielectric nanostructure to enhance the probability of dipole-forbidden atomic transitions has been proposed in \cite{Klimov1996}, where enhanced radiation emission of an atom placed next to a dielectric sphere has been considered and the subject is continuously topical in solid state systems \cite{Deng,Sain}. The evanescent fields also can be used for such enhancement, as demonstrated experimentally in \cite{Tojo2004}. Yet another way to achieve an enhancement of quadrupole transition probability is to use polarized standing waves \cite{Yang2010}. Among various field focusing techniques, one of the most effective ones is the use of surface plasmons; they have been proposed as a means to enhance both atomic \cite{Deguchi2009,Kern2012} and excitonic \cite{Okuda2006,Neubauer} transitions. 

In this paper, we focus on excitons in Cu$_2$O and their interaction with surface plasmons. The aim to provide some initial estimations of the enhancement of specific excitonic transitions in a few selected copper-based plasmonic systems. We start from low $n$ excitons and then illustrate some general tendencies relevant to highly excited Rydberg states.

\section{Copper plasmons}
A surface plasmon is an electromagnetic excitation forming on the interface between two media characterized with opposite signs of dielectric permittivity at some specified frequency. Usually, the medium with negative permittivity is a metal and the other one is any dielectric. Let's assume that the permittivities of these two materials are $\epsilon_1(\omega)<0$ (a metal) and $\epsilon_2(\omega)>0$ (a dielectric). In such a system an electromagnetic wave mode bounds to the material interface can propagate. This mode is characterized by a wave vector \cite{Chubchev}
\begin{equation}\label{dysp_plazmon}
k(\omega)=k_0\sqrt{\frac{\epsilon_1\epsilon_2}{\epsilon_1+\epsilon_2}},
\end{equation}
where $k_0=\frac{\omega}{c}$ is the free space wave vector and $c$ is the speed of light in vacuum.  Propagating plasmon modes can be excited when $\epsilon_1\epsilon_2<0$ and $\epsilon_1+\epsilon_2<0$, so that $k$ is a real number \cite{Chubchev}.

Here we  concentrate on the visible part of the spectrum where absorption  of yellow exciton series in Cu$_2$O are observed. This region corresponds to wavelength $\lambda \approx 570$ nm. Copper, while being an uncommon choice for plasmonic applications, is particularly well suited for this spectral region \cite{copper_plasm}. The permittivity of copper at $\lambda \approx 570$ nm is \cite{Hollstein} $\epsilon_1 \approx -7.5 + 1.77i$ (see Appendix B). The absolute value of the real part of permittivity is roughly the same as the one characterizing copper oxide; $\epsilon_2 \approx 7.5$. This means that not only surface plasmons can be readily excited on the Cu-Cu$_2$O interface, but also that they will be characterized by a large value of $k$ since $\epsilon_1+\epsilon_2\rightarrow 0$, which stands for that highly focused, short wavelength plasmon modes can be excited. In particular, for the material data specified above, $Re~k \approx 4.5 k_0$, which implies that the plasmon wavelength is reduced by a factor of 4.5  compared to the free space wavelength.

The plurality of Rydberg excitonic states in cooper oxide offers the wealth of available excitonic states in a wide range of frequencies. Selection rules determine whether an optical transition is forbidden or allowed, which  depends on symmetry properties of states involved in such a transition.  In Cu$_2$O the valence band is characterized by an even symmetry therefore  $S$ and $D$ excitons are dipole-forbidden but quadrupole allowed.

\begin{figure}[ht!]
\includegraphics[width=.7\linewidth]{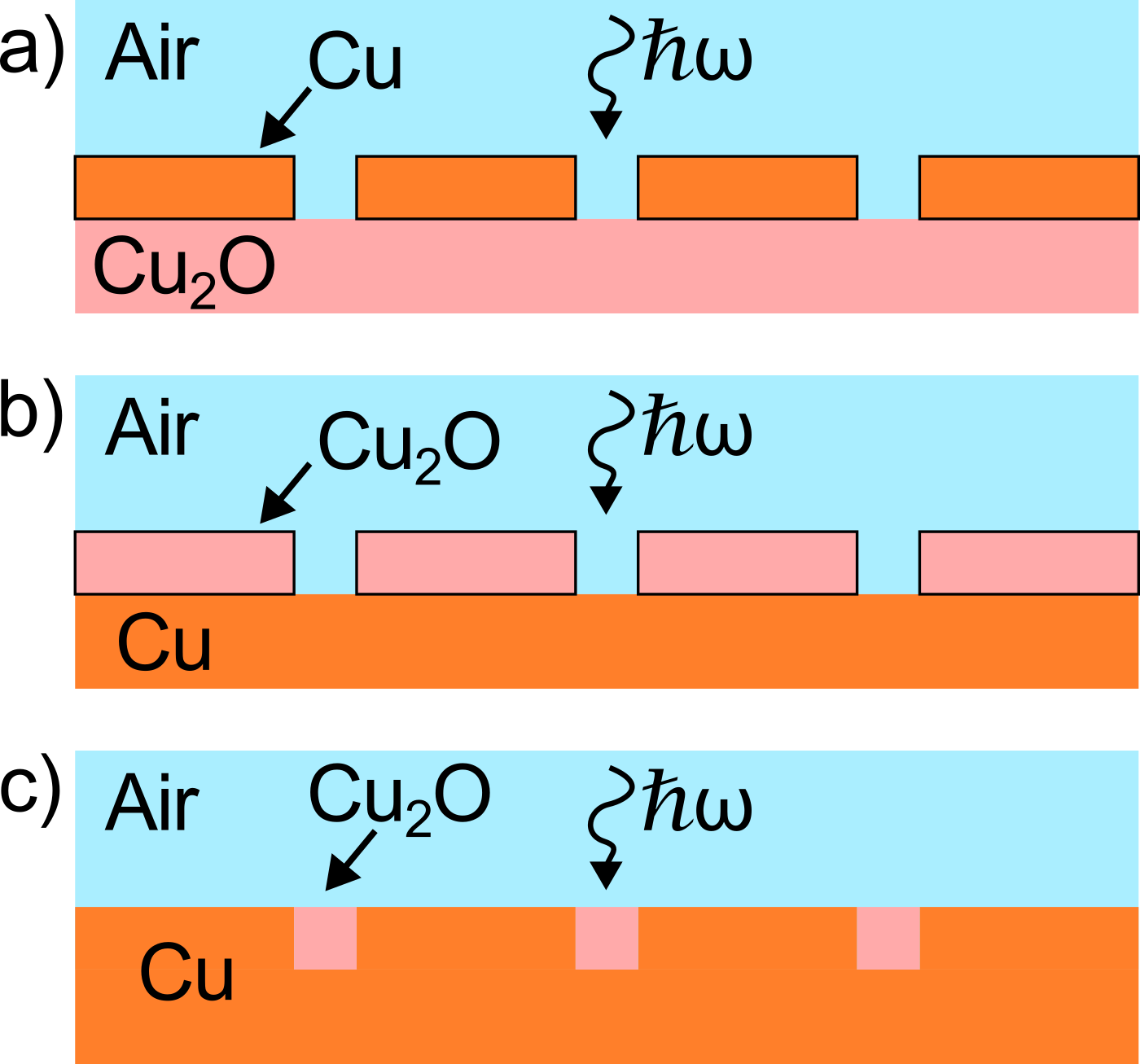}
\caption{Two-dimensional cross-sections of three types of Cu-Cu$_2$O nanostructures.}\label{fig:struktury}
\end{figure}

As it will be discussed in the next section, in order to amplify the probability of quadrupole exciton transition, large electric field gradient should be created. By compressing an electromagnetic wave traveling in $z$ direction by a factor of 4.5, one gets an increase of $\partial E/\partial z$ by the same factor. Even disregarding any effects originating from the geometry of the plasmonic structure \cite{Chan2007}, local field gradient is enhanced substantially by plasmons.
By fine-tuning the geometrical parameters of nanostructures the plasmonic resonaces can be aligned with the optical properties of excitons in Cu$_2$O;  in order to  confine the field even further, three types of structures depicted on Fig. (\ref{fig:struktury}) will be considered in this paper.  

The first system, taken from Ref. \cite{Okuda2006}, consists of a copper layer deposited on top of copper oxide. The layer contains a periodic set of $\sim 100$ nm slits that allow the illuminating light (top) to enter Cu$_2$O. Surface plasmons form on the Cu-Cu$_2$O interface, with particular emphasis on the vicinity of the slits. 
 
The second structure consists of continuous Cu layer with nm to $\mu$m sized copper oxide nanoparticles deposited on top of it. This geometry is proposed in order to maximize the quadrupole transition amplification by restricting the possible location of Cu$_2$O excitons to a small volume.

Finally, the third setup consists of small cavities filled with Cu$_2$O, embedded in a comb-like copper structure. This geometry has potential to confine both the excitons and the plasmon field to the volume even smaller than nanoparticles.

\section{Enhancement of quadrupole transitions}

In this section, the theoretical description of the enhancement of the quadrupole transition is presented and approximate solution for larger excitons is proposed.

\subsection{Point quadrupole model}
Let's assume we have a single exciton located at $\vec{r}=0$. An external electric field $E$ surrounding the exciton can be expanded with Taylor series; by keeping the first two terms, one obtains
\begin{equation}
\vec{E}=\vec{E}(0)+\vec{r}\cdot(\nabla \vec{E}),
\end{equation}
where $(\nabla \vec{E})=\left[\frac{\partial E_i}{\partial r_j}\right]_{i,j}$ is Jacobian matrix of $\vec{E}$. The potential $\varphi = -\int \vec{E} \cdot d\vec{r}$ will be given by
\begin{eqnarray}
&&-\varphi = \vec{E}(0)\int d\vec{r} + \int\vec{r}\cdot(\nabla \vec{E})\cdot \vec{dr},
\end{eqnarray}
which yields \cite{Barron1973}
\begin{equation}
\varphi = \phi(0)-\vec{E}(0)\cdot \vec{r} - \frac{1}{2}\sum\limits_{i,j}r_ir_j\frac{\partial E_j}{\partial r_i}
\end{equation}
where constant part of potential $\phi(0)$ is added for completeness \cite{Neubauer}. With a charge $q$ (for the case of exciting S exciton, $q$ is a charge of valence band electron), the energy is  
\begin{equation}\label{eq:energy}
W=q\varphi(0)-\vec{E}(0)\cdot \vec{p} - \frac{1}{6}\sum\limits_{i,j}Q_{ij}\frac{\partial E_j}{\partial r_i}
\end{equation}
with dipole moment $\vec{p}=q\vec{r}$ and quadrupole moment $Q_{ij}=q(3r_ir_j-r^2\delta_{ij})$ \cite{Jackson}.
The interacting part of hamiltonian of such a system reads
\begin{equation}
V=-\vec{p} \cdot \vec{E}(0) - Q : (\nabla \vec{E})
\end{equation}
where $A:B=\sum\limits_{i,j}A_{ij}B_{ji}$ \cite{Kern2012}; the above hamiltonian is then used to calculate the transition rate $\gamma$, which can be estimated from the Fermi's golden rule
\begin{equation}\label{eq:gamma}
\gamma \sim \left|\langle\Psi_2|V|\Psi_1\rangle\right|^2,
\end{equation}
where $\Psi_1$ and $\Psi_2$ are initial and final state of the system. The key property of the above relation is the dependence of $\gamma \sim (\nabla E)^2$; plasmonic systems, as described in last section, are characterized by highly focused field corresponding to large values of $(\nabla E)$ compared to free space waves.

Recently Neubauer \emph{et al} observed \cite{Neubauer} that the confined field of surface plasmons can enhance the dipole-forbidden transition from valence band to even exciton states ($S,D$...), so that they can become visible in absorption spectrum. In other words, the usually negligible oscillator strengths of these transitions  increase   reaching values comparable to these  characterizing dipole transitions. In the context of the yellow exciton series in Cu$_2$O that are the focus of this paper, the goal is to observe $S$ excitons in the spectrum that is normally dominated by spectral lines corresponding to $P$ excitons.

With the help of excitation probability between states  $\Psi_1$ and $\Psi_2$,  $P(r)\sim |\langle \Psi_2|Q|\Psi_1 \rangle \nabla E|^2$, one can calculate the enhancement factor \cite{Okuda2006,Deguchi2009}
\begin{eqnarray}\label{Okuda_1}
&&\frac{\gamma}{\gamma_0}=\frac{(n+1)^2}{4E^2n^2k_0^2}\left[\left(\frac{\partial E_y}{\partial z}+\frac{\partial E_z}{\partial y}\right)^2+\left(\frac{\partial E_z}{\partial x}+\frac{\partial E_x}{\partial z}\right)^2\right.\nonumber\\&&+\left.\left(\frac{\partial E_x}{\partial y}+\frac{\partial E_y}{\partial x}\right)^2\right],
\end{eqnarray}
where $nk_0$ is its wave vector and $n$ is the refraction index of the medium. 

The excitation of 1$S$ state in Cu$_2$O has been discussed  in gold slit array  \cite{Okuda2006}, finding amplification factor of the order of 10-100. However, there's an open question of how large the amplification can get for higher excitonic states. This issue is addressed in section below.

\subsection{Higher excitonic states}
An important assumption made in the above derivations is that the Eq. (\ref{eq:energy}) is valid for immediate vicinity of $\vec{r}=0$ and is most accurate for excitons that are much smaller than the structure they interact with. For example, in \cite{Kern2012} interaction between cesium atom (under 1 nm diameter) and 50 nm nanoantenna is considered; similarly, Ref. \cite{Okuda2006} is devoted to excitation of 1$S$ exciton in Cu$_2$O, with a diameter of approximately 2 nm. On the other hand, the Rydberg excitons observed experimentally in Ref. \cite{Neubauer} are characterized by quantum number $n \sim 6$. Here in particular, we will consider interactions with $1S$, $2S$ and $3S$ excitons, characterized by a radius of 1.1, 4.4, 9.9 nm correspondingly.

As discussed in \cite{Neuman}, there are many systems where plasmonic field is confined to such a degree that systems such as quantum dots, organic molecules or Rydberg excitons are no longer can be considered as  point dipoles/quadrupoles. It is known that many quantum processes such as spontaneous emission become nonlocal in such conditions \cite{Stobbe}. Specifically for Cu$_2$O, it has been experimentally demonstrated that nonlocal interactions become significant when exciton radius $r_n$ becomes similar to the wavelength of incident light; the match between the size of exciton wave function governed by Schr\"{o}dinger’s equation and electromagnetic wave described by Maxwell’s equations becomes a significant factor \cite{Takahata2018}; in particular, the emission intensity from a thin Cu$_2$O layer is shown to be dependent on the layer thickness \cite{Takahata2018}. Surface plasmons, with their reduced wavelength, can lead to the nonlocal effects even for relatively low excitonic states. 

Nonlocal interactions between plasmons and quantum emitters are a dynamically developing field, with multiple theoretical approaches and approximations being proposed\cite{Eriksen2024}. In the context of the systems presented here, it is useful to mention the approach proposed in \cite{Neuman} that considers an extended radiation source in the form of large molecule. By calculating the spatial distribution $\rho(\vec{r})$ of the oscillating electric charge over a molecule emitter, one can estimate the transition density between two states $i,j$ $\rho_{ij}(\vec r)=e\langle j|\rho(\vec r)|i\rangle$. Finally, the obtained transition density can be used to calculate energy
\begin{equation}\label{eq:neu}
E = \int\limits_{V}\rho_{ij}(\vec r)\varphi(\vec r)d^3r,
\end{equation}
integrating over the volume of the emitter. 

In the case of excitons, similar calculation could be performed; the ground and excited states are described by Bloch functions of the valence band electron \cite{Heck017} and hydrogenlike excitonic wavefunction, respectively. A full \emph{ab initio} calculation of the dipole and quadrupole oscillator strengths is complicated, especially for odd parity excitons \cite{Schweiner2017}. However, what we are interested in this paper is the amplification of these transitions and not their absolute value. To that end, we note that the key quantities that need to be evaluated are the spatial derivatives of the field surrounding the plasmonic structure in the context of finite sized excitons. We aim to provide a simple heuristic approach that goes beyond the point quadrupole expression in Eq. (\ref{Okuda_1}), taking into account several phenomena that can be responsible for lower than expected amplification of quadrupole transitions observed in \cite{Neubauer}. In particular, we propose a two stage averaging procedure as outlined below.

\subsection{Field averaging approach}
First, we note that the electric field distribution around nanostructure, obtained with numerical methods such as FDTD simulation employed here, has a finite resolution. Thus, the spatial field derivatives have to be approximated by a fraction $\Delta E/\Delta x$, where $\Delta E$ is the difference between field values in two neighboring grid points and $\Delta x$ is the spatial step of the simulation. While such an approach provides some challenges in resolving highly confined field of plasmonic structures \cite{Okuda2006}, it can also serve as an useful tool for modeling finite sized dipoles and quadrupoles. Specifically, in a case of a quadrupole transition of finite-sized exciton, one can evaluate the mean field gradient within exciton volume by considering a value of $\Delta x$ that is comparable to the exciton diameter. This serves as a good first approximation of the integration over charge density distribution in Eq. (\ref{eq:neu}). In particular, this type of averaging greatly reduces the influence of local field maxima that are much smaller than the exciton; indeed, the necessity to match the size of the plasmonic structure to the exciton diameter is noted in \cite{Neubauer}. At the same time, for an efficient excitation of surface plasmons, one needs to match the nanostructure size to the wavelength of incident light, which is on the order of $\lambda=600$ nm. Thus, geometric features with size $\lambda/2$, $\lambda/4$ etc. are beneficial.

A second important aspect of the plasmon-exciton interaction in a nanostructure is the uncertain location of excitons. In the case where excitons are created with a free wave propagating in Cu$_2$O crystal, the excitons are located in random positions, with a distribution that is approximately uniform in small scale and follows Lambert-Beer law in a larger scale due to the fact that exciton density is proportional to the absorbed power density. However, in the immediate vicinity of the metallic structure, the energy of excitons is altered due to confinement effects, so that it no longer matches the photon energy of incident field. This prevents effective creation of excitons by incident field. Therefore, a ,,dead zone'' with a thickness comparable to exciton radius is formed around the plasmonic structure.  This is an important effect that potentially greatly reduces the possible amplification due to the fact that excitons cannot be created in the area where the field gradient is the largest.

To sum up, we propose a two-stage field averaging procedure:
\begin{itemize}
\item The individual gradients $\partial E/\partial x$, $\partial E/\partial y$ etc. are replaced with fractions containing discrete spatial steps  $\Delta x$, $\Delta y$ proportional to exciton radius.
\item The amplification factor $\eta(\vec{r})=\gamma/\gamma_0$ is averaged over the volume surrounding the nanostructure, with a given exciton probability density $\rho_{exc}(\vec{r})$, yielding average value $\hat{\eta}$. 
\end{itemize}
The averaged-out amplification factor is given by
\begin{equation}
\hat{\eta}=\frac{\int\rho_{exc}(\vec{r})\eta(\vec{r})d^3r}{\int\rho_{exc}(\vec{r})d^3r},
\end{equation}
integrating over volume taken by Cu$_2$O in which the exciton density $\rho_{exc}$ is nonzero.

\section{Amplification of quadrupole transition by plasmonic structures}
In this section, we apply the theoretical description of quadrupole transition amplification to the electromagnetic field generated by selected plasmonic structures, starting with a one-dimensional model.

\subsection{One-dimensional model}
Before analyzing full nanostructures, let's consider a single metal-dielectric interface. Surface plasmons forming on such an interface are characterized by a real wave vector component parallel to the interface, given by Eq. \ref{dysp_plazmon}. In the direction perpendicular to the surface, the wave vector component is given by
\begin{equation}
k_\perp=\frac{\omega}{c}\sqrt{\frac{\epsilon_{1,2}^2}{\epsilon_1+\epsilon_2}},
\end{equation}
where $\epsilon_{1,2}$ are the permittivities of the dielectric and metal. Since $\epsilon_2<0$ and $|\epsilon_2|>|\epsilon_1|$, the $k_\perp$ is imaginary and describes evanescent that decay exponentially away from the interface. Notably, $k_\perp$ is highly dependent on the sum $\epsilon_1+\epsilon_2$; this means that depending on the choice of metal and frequency on the plasmon, one can achieve a plasmon field that is either highly focused or extends over long range. This is particularly relevant to the realization of plasmon-exciton coupling, where the minimum distance between metal surface and exciton is limited by exciton's size.

\begin{figure}[ht!]
\includegraphics[width=.9\linewidth]{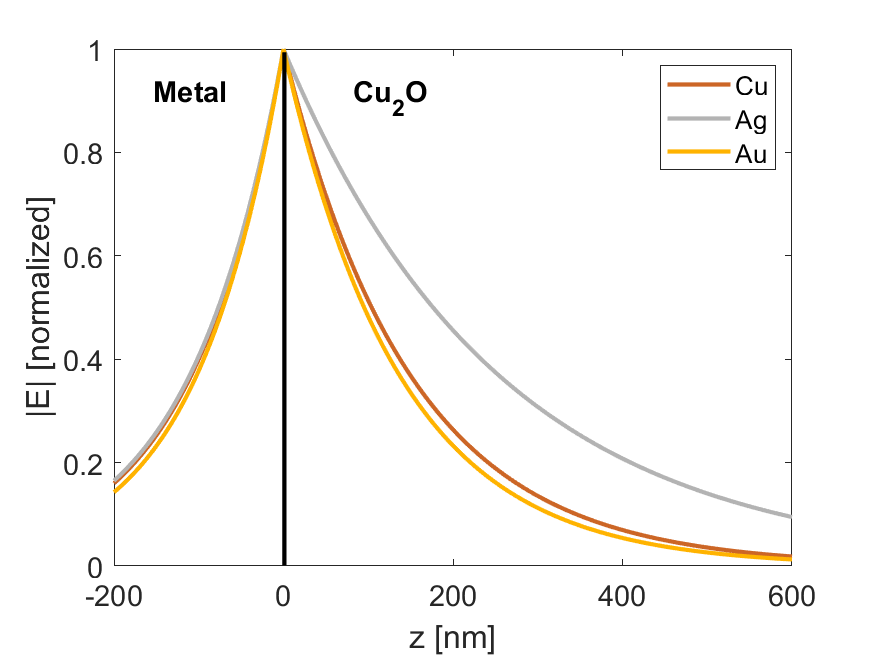}
\includegraphics[width=.9\linewidth]{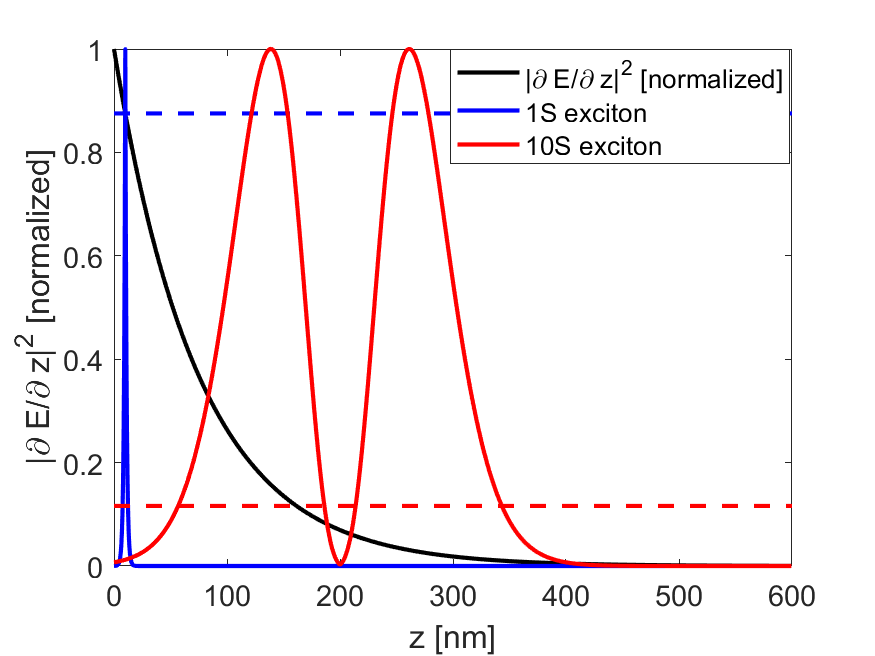}
\caption{a) Electric field on the metal-Cu2O interface; b) Field gradient and two exciton probability densities for reference.}\label{fig:decay}
\end{figure}

Let's consider a specific example of three metals typically used in plasmonics (copper, silver, gold), in a frequency range where excitonic resonances are located. The material properties are discussed in Appendix A. The exponentially decaying field of surface plasmons is shown on Fig. \ref{fig:decay} a). The frequency is set to match the energy of 1S exciton. 
One can see that differences between copper and gold are quite minor. However silver, due to its larger absolute value of permittivity, is characterized by a slower field decay, with 10$\%$ of the initial field reached at a distance of $\sim 600$ nm from the interface. This makes it particularly suitable for designs involving large Rydberg excitons. 

As mentioned before, the exciton radius is given by $r_n=a_b n^2$, with $a_b=1.1$ nm. A metal surface placed next to the exciton acts as a quasi-infinite potential barrier, causing the exciton to enter a semi-confined state resembling a half-cavity. Such a half-cavity will cause a significant energy shift of the excitonic state when the distance between an exciton and a metal surface is smaller than an exciton radius. Thus, $r_n$ can be assumed to be the minimum possible distance between an  exciton and a metal interface. However, it should be stressed that in the case where the excitons are created with the plasmon field, the exciting field
is very wide spectrally, allowing for excitation of energy-shifted states.

Let's consider  $1S$ and $10S$ excitons placed near a copper surface, at a minimum allowed distance from the metal-dielectric interface. The effective size of the exciton is determined by the probability density $|\Psi|^2$, where $|\Psi|$ is the hydrogenlike wavefunction of the exciton state. On Fig. \ref{fig:decay} b), one can see that the $1S$ exciton is effectively point-like and can be excited very close to the interface. The $10S$ state, on the other hand, is characterized by effective radius of $r \sim 100$ nm. As described in the field averaging section IIIC, when exciton cannot be treated as a point particle, the field gradient can be calculated as an average over the volume taken by exciton. In the one-dimensional model considered here, the amplification of the quadrupole transition is obtained from Eq. (\ref{Okuda_1}) by considering only $\partial E/\partial z$ terms. By averaging them over the area where exciton probability density is nonzero, one obtains values marked on Fig. \ref{fig:decay} b) by dashed lines. The difference between $1S$ and $10S$ state is quite considerable, illustrating how the lowest exciton states have an advantage in possible amplification. However, since quadrupole transition moment scales with exciton radius, the optimization of emission energy is a nontrivial problem. 

At this point, it is useful to briefly discuss some issues related to dipole emission, as opposed to quadrupole emission considered here. The reflective metal surface modifies the properties of a quantum emitter (such as exciton), affecting both the emission rate and frequency \cite{Scerri2017}. In particular, one can analyze this problem with the method of images, where a fictious dipole is placed inside the metal layer. In such a case, a dipole-dipole interaction occurs between exciton and its image. As a result, in plasmonic systems involving dipole transitions of $P$ excitons, one has to replace the exciton radius with considerably larger Rydberg blockade radius when estimating the minimum distance between exciton and metal layer. Here, for $S$ excitons, only a quadrupole-quadrupole interaction occurs, that is very short range $\sim 1/R^5$ and negligible unless very high principal quantum number states are considered.

\subsection{Metallic grating}
\begin{figure}[ht!]
\includegraphics[width=.95\linewidth]{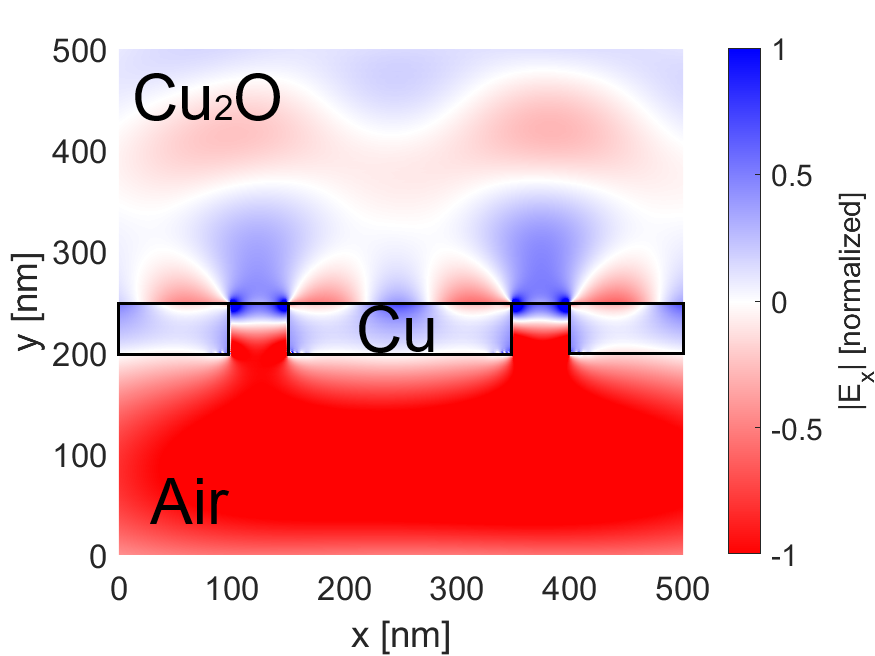}
\caption{Spatial distribution of electric field in a grating system, obtained from FDTD simulation.}\label{fig:oku1}
\end{figure}

First, let's consider a system proposed in Ref. \cite{Okuda2006}, where a metallic grating is deposited on the surface of Cu$_2$O crystal, as depicted on Fig. \ref{fig:oku1}. In contrast to \cite{Okuda2006}, we use copper for the grating material, similarly to our earlier works \cite{copper_plasm,konwerter}. 

 Fig. \ref{fig:oku1} shows a two-dimensional cross-section of the system located on $xy$ plane. Three cases are considered, where the incident field of a wavelength $\lambda \sim 600$ nm is tuned specifically to the $1S$, $2S$ or $3S$ exciton line. The field illuminates the metallic grating from the bottom, exciting surface plasmons on the grating structure and also entering Cu$_2$O through the 40 nm gaps in the grating. Inside copper oxide, the wavelength is reduced to $\lambda'=\lambda/\sqrt{7.5} \approx 220$ nm. However, at the Cu-Cu$_2$O interface, excitation of surface plasmons leads to the formation of highly localized field maxima with a size of 50 nm or less. This deep subwavelength field concentration becomes a source of a large local increase of $\nabla E$ and consequently large potential amplification.  

\begin{figure}[ht!]
\includegraphics[width=.95\linewidth]{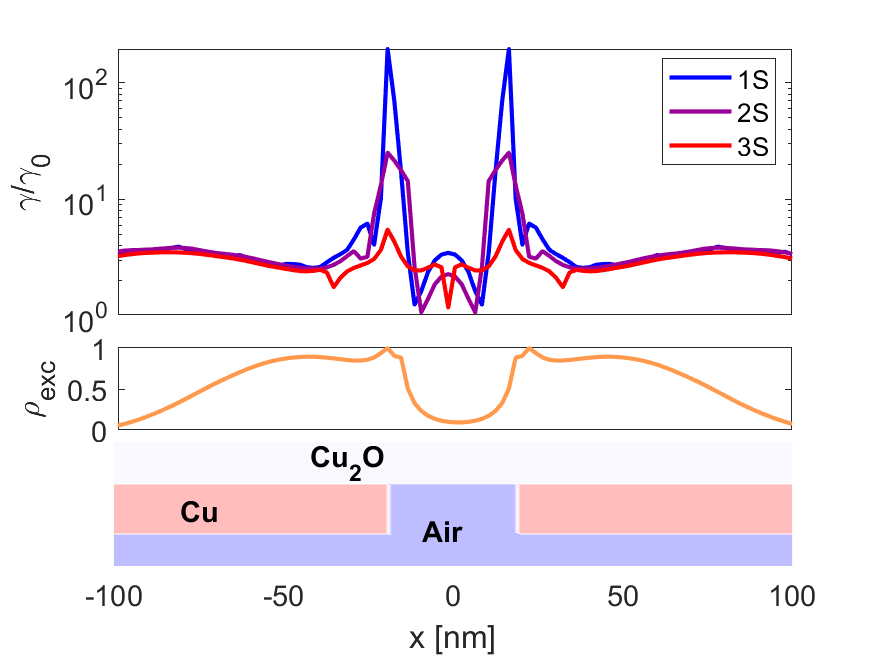}
\caption{Fragment of the system  showing one slit (bottom panel), local density of excitons (middle panel) and quadrupole transition amplification (top panel).}\label{fig:oku1b}
\end{figure}

Let's investigate the quadrupole transition amplification taking place in the neighborhood of a single slit. In particular, the enhancement of the $1S-3S$ exciton excitation is calculated from Eq. (\ref{Okuda_1}), taking into account the two above-mentioned averaging procedures. The considered system is shown on the bottom of Fig. \ref{fig:oku1b}.

The top panel shows the amplification factor $\eta=\gamma/\gamma_0$ calculated for the area immediately above the Cu-Cu$_2$O interface (y=250 nm on Fig. \ref{fig:oku1}). There are two sharp peaks corresponding to the edges of the slit (Fig. \ref{fig:oku1b}, bottom panel), where local maxima of the field are located (see also Fig. \ref{fig:oku1}). The height of these peaks depends strongly on the exciton radius $r \sim n^2$, indicating that the principal quantum number $n$ of the excitonic state has a huge impact on the transition amplification, with smaller excitons having an advantage. 

As the exciton gets larger, the local field gradient is averaged over larger volume, leading to a smaller maximum value. Furthermore, due to the potential barrier between Cu$_2$O and Cu, the exciton cannot be located at a distance from the metal surface that is smaller than $r$. Therefore, larger excitons are located further from metal surface, where local field intensity is smaller. Our theoretical result gives support to the observation in \cite{Neubauer}, where the amplification of the oscillator strength of $6S$ state was smaller than $5S$ state.

In the case of $1S$ exciton, characterized by $r=1.1$ nm, the local amplification reaches a value of $\eta \approx 140$. However, as mentioned before, there is no way to guarantee that exciton will be located exactly next to the Cu-Cu$_2$O interface. Instead, one has to consider the density of excitons in the entire Cu$_2$O layer, which is proportional to electromagnetic field energy density $\rho_{exc} \sim \epsilon E^2$. The normalized exciton density is shown on Fig. \ref{fig:oku1b}, middle panel. While it reaches the largest value at the amplification peaks, its value is nonnegligible in other regions as well. In particular, the entire Cu-Cu$_2$O interface supports surface plasmons with considerable field intensity $E$ but smoother field distribution (and thus smaller field gradient) than the slit edges. Thus, one obtains averaged-out amplification factor that is considerably lower than the peak value. In particular, for $1S$ exciton one obtains peak value of the amplification factor $\eta \approx 140$ and average value $\hat{\eta} \approx 17$. 

\begin{figure}[ht!]
a)\includegraphics[width=.8\linewidth]{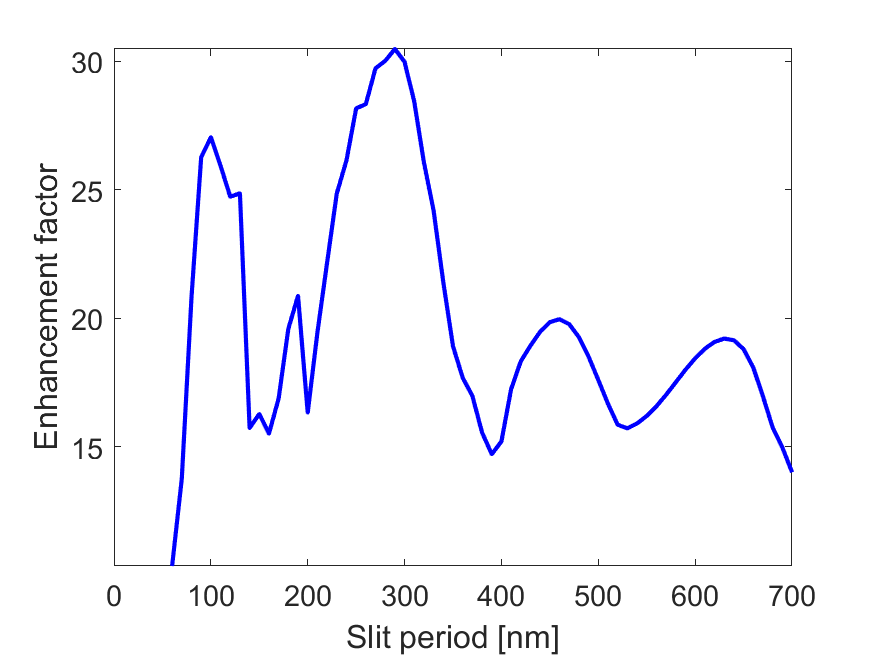}
b)\includegraphics[width=.8\linewidth]{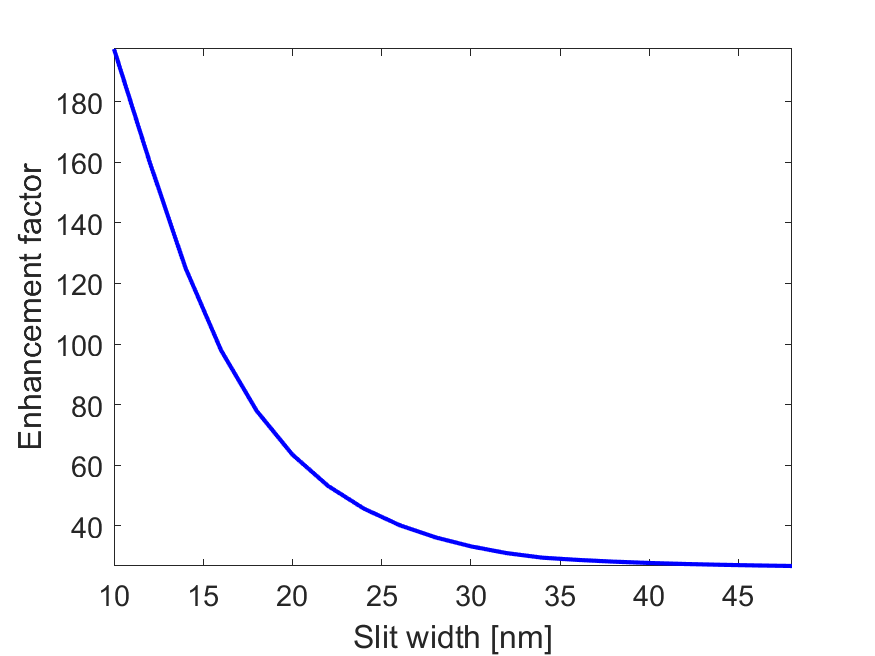}
\caption{The quadrupole transition amplification as a function of a) slit period and b) slit width.}\label{fig:oku2}
\end{figure}

The two geometric parameters describing the grating is its period and slit width. Thus, the quadrupole amplification factor $\gamma/\gamma_0$ has been calculated as a function of these parameters. The results are shown on Fig. \ref{fig:oku2}.

The dependence of amplification on slit period, shown on Fig. \ref{fig:oku2} a), is consistent with the results presented in \cite{Okuda2006}, confirming the validity  of the theoretical and numerical calculations and indicating that the general properties of the plasmonic system are consistent regardless of the choice of the metal used to fabricate the structure. Maximum enhancement of the transition is of the order of 30. There are several maxima of amplification corresponding to the period of 110, 190, 290, 450 and 650 nm. These values are close to $\lambda'/2$, $\lambda'$, $3\lambda'/2$, $2\lambda'$ and $3\lambda'$, correspondingly, indicating that standing wave plasmon modes are excited in the system. In order to reach a high amplification, strong local maxima of field are needed; these are facilitated by standing wave patterns in the field, leading to the even fractions of wavelength being optimal. Naturally, the exact optimal size of nanostructure will depend on its shape, but the general conclusion is that the nanostructure size it tied to the wavelength of the light which, in turn, is tuned to the exciton energy.

The dependence of amplification on slit width is more straightforward; as shown on Fig. \ref{fig:oku2} b), the enhancement is approximately inversely proportional to the gap width. This is also consistent with Ref. \cite{Okuda2006} and similar to the case of various other plasmonic nanostructures employing very narrow gaps for large field concentration \cite{Bahari2016,Min2022}. At this point, it is important to discuss some limitations of the presented model. It is expected to correctly predict the system properties for a small exciton; in particular, results for $1S$ exciton are consistent with \cite{Okuda2006}). However, for large excitons, including Rydberg excitons, the proposed averaging scheme serves only as a rough approximation of the nonlocal processes taking place in the system. As mentioned in \cite{Min2022}, metallic slit like the one described here can sustain plasmonic oscillations producing both dipole and quadrupole radiation patterns; it is expected that the latter type is beneficial to the amplification of quadrupole excitonic transitions. Similarly, in \cite{Bahari2016} it is shown that nanocube antenna with 4-way symmetry is optimal for exciting quadrupole radiation modes. Therefore, there is an interplay between  an exciton position, size and nanostructure size that goes beyond simple averaging proposed here. However, regardless of the approach, it is evident that a fine control over an exciton location within nanostructure would be beneficial for obtaining optimal amplification. This issue, also raised in \cite{Neubauer}, is addressed in the next considered system.

\subsection{Cu$_2$O nanoparticles}

After analyzing the plasmonic system proposed in Ref. \cite{Okuda2006}, verifying the our approach\textit{/averaging procedure?} and extending the results to larger exciton states, one can propose some alternative nanostructures that can be used to maximize the transition amplification.

\begin{figure}[ht!]
\includegraphics[width=.95\linewidth]{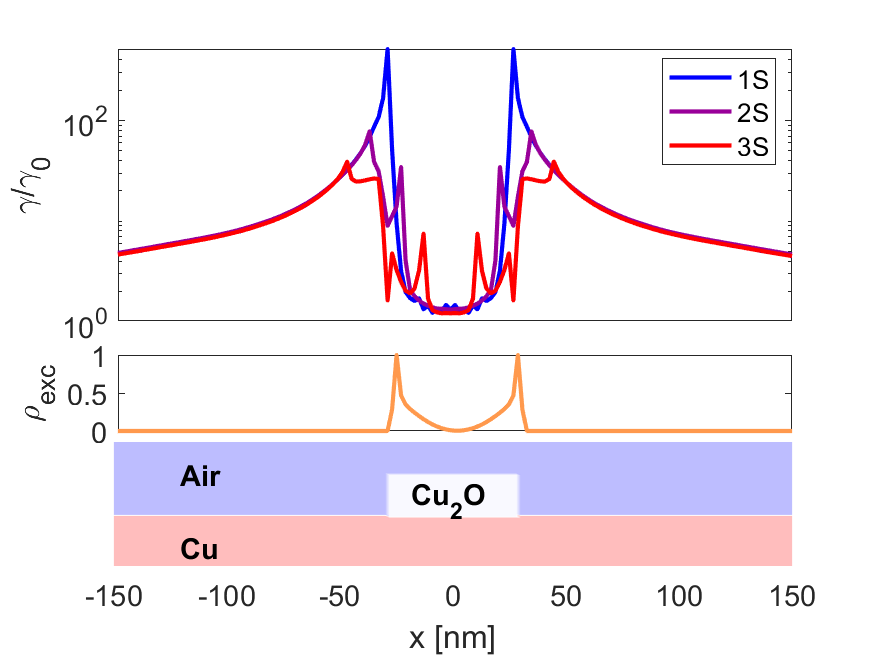}
\caption{Fragment of the system  showing one nanoparticle (bottom panel), local density of excitons (middle panel) and quadrupole transition amplification (top panel).}\label{fig:nano1}
\end{figure}

Let's consider a system where small Cu$_2$O nanoparticles are deposited on Cu surface. Schematic of the system is shown on Fig. \ref{fig:nano1}.

The system is illuminated from the top. An incident wave is mostly reflected from the copper surface, but a part of its energy is used to excite plasmons on Cu-Cu$_2$O interface. Like as in the previous system, the strongest field concentration, and thus the largest amplification, corresponds to structure edges. Here, the peak value is $\eta \approx 450$ (Fig. \ref{fig:nano1}, top panel). The large advantage is that the excitons can only be created within small Cu$_2$O nanoparticle. Therefore, the region where $\rho_{exc}$ is nonzero is small (middle panel), leading to a large average amplification $\hat{\eta} \approx 105$ for $1S$ exciton, which improves upon the previous considered structure by a factor of $\sim 6$. It should be noted that Cu$_2$O nanoparticle is represented by a 50 nm x 20 nm rectangle, but a less regular shape would work as well. Importantly, while the obtained amplification is larger, the number of excitons in the system is much smaller than in the previous example. Thus, this setup is more suited for performing operations on a very low number of excitons, possibly a single Rydberg exciton. Like in the previous system, the amplification of transition to higher excitonic states is suppressed by the averaging; for $3S$ state it is already $\sim 10$ times smaller than for $1S$ state, with peak value of the order of 40.

\subsection{Comb structure}
\begin{figure}[ht!]
\includegraphics[width=.95\linewidth]{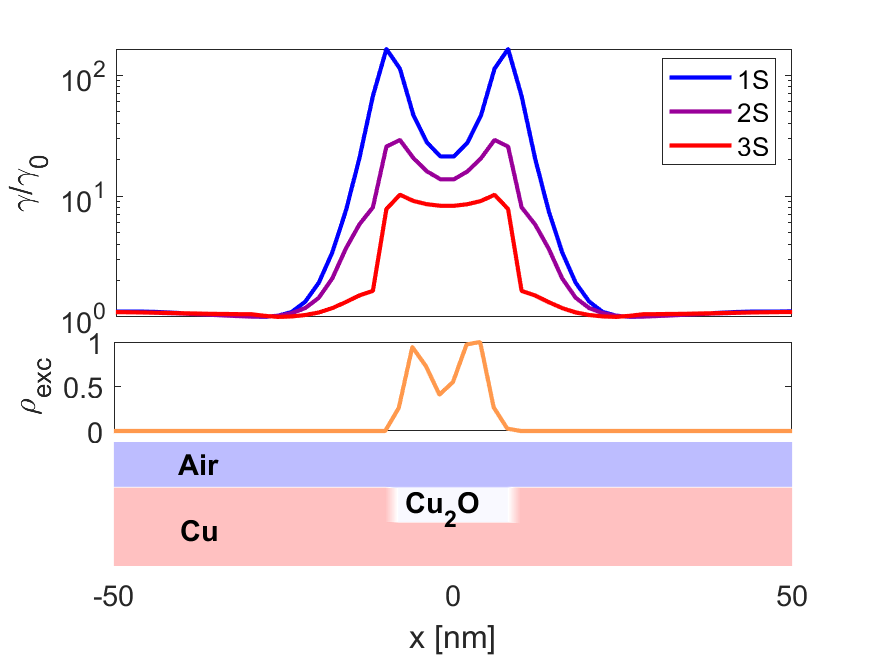}
\caption{Fragment of the system showing one valley (bottom panel), local density of excitons (middle panel) and quadrupole transition amplification (top panel).}\label{fig:comb1}
\end{figure}

The third plasmonic system considered in this paper consists of a comb-like metallic structure with narrow slits filled by Cu$_2$O. On the Fig. \ref{fig:comb1} bottom panel, one such slit is depicted; it has a width of 20 nm and depth of 10 nm. 

Similar setup has been discussed in our paper \cite{konwerter} in a context of frequency conversion with Rydberg excitons; it has been shown that microwave fields can be highly confined in the slits, leading to small effective mode volume and large amplification of microwave transition probability by a factor of $\sim 10^4$. In the optical regime considered here, the much shorter wavelength plasmonic modes cannot be confined to such a degree, leading to a smaller amplification $\eta \sim 160$, which is of the same order as in the other discussed systems. Similarly to the case of nanoparticle, the volume filled with copper oxide is very small, limiting the possible locations of excitons to the area where large amplification is achieved. Thus, the average amplification is $\hat{\eta} \approx 75$.

\subsection{Application to other quadrupole transitions}

In contrast to the transition between ground state and $S$ exciton state considered here, various transitions between two excitonic states are also possible. Due to the small energy spacing between high $n$ levels, the absorption/emission spectrum of these transitions is in the microwave region \cite{maser} and they can also be amplified with plasmonic structures \cite{konwerter}. In particular, quadrupole transitions between $S$ and $D$ Rydberg atom states have been discussed in \cite{Deguchi2009}, demonstrating $10^2$ enhancement with a metallic grating. Like in the examples above, it is expected that for relatively large excitons, enhancement would be reduced. As mentioned before, for a principal quantum number $n$, the exciton radius (and thus, the minimum distance to the metal surface) scales as $n^2$. The exponentially decaying field gradient on the metal-dielectric interface is thus $\sim \exp(-r_n)$. However, for $nS \rightarrow nD$ transition, the quadrupole transition moment will be proportional to $r_n^4$. Therefore, the total strength of the transition will be given by an expression of the type $\sim r_n^4 \exp(-2r_n)$, with a local maximum at some specific exciton size dependent on multiple factors such as plasmonic structure geometry. Therefore, optimization of this type of systems could be highly nontrivial.

\section{Conclusions}
The potential of the quadrupole transition amplification of several types of plasmonic structures has been investigated, with particular emphasis on the excitation of $S$ excitons in Cu$_2$O. Building upon available theoretical work describing an interaction between plasmons and point quadrupoles, an approximation has been developed in order to calculate the amplification factor for larger exciton states. After establishing the general scaling of the phenomenon with the principal quantum number $n$, two types of plasmonic structures have been proposed in order to maximize the plasmonic amplification for Cu$_2$O excitons specifically. The obtained results shed some light on the recent experimental data \cite{Neubauer} and could pave a way to further application of plasmonic structures to highly excited Rydberg states.

\section{Appendix A: FDTD method}

The Finite-Difference Time-Domain (FDTD) method \cite{Yee} has been used to obtain the spatial field distributions of surface plasmons excited in nanostructures considered in this paper. Due to the fact that this numerical approach is based directly on Maxwell's equations, it provides high accuracy of predictions regardless of structure complexity, with no simplifying assumptions. 

Here, similarly to \cite{Okuda2006}, we take advantage of the structure symmetry and use a two-dimensional calculation scheme. The analyzed system is a cross-section of the structure placed on $xy$ plane. The structure is assumed to be large (much larger than wavelength) in $z$ direction. The computational domain consists of discrete set of points placed on a rectangular grid, with individual points located $\Delta x = 1$ nm apart. At every point, a new value of electric and magnetic fields $\vec{E}(x,y,t+1)$, $\vec{H}(x,y,t+1)$ are calculated based on the previous values $\vec{E}(x,y,t)$, $\vec{H}(x,y,t)$, according to the evolution equations
\begin{eqnarray}\label{FDTD_finalne}
&&\frac{\partial E_y(x,y,t)}{\partial x}-\frac{\partial E_x(x,y,t)}{\partial y}=-\mu_0\frac{\partial H_z(x,y,t)}{\partial t},\\
&&\frac{\partial H_z(x,y,t)}{\partial y}=j_{x}(x,y,t)+\epsilon_0\frac{\partial E_x(x,y,t)}{\partial t} -\frac{\partial P_x(x,y,t)}{\partial t},\nonumber\\
&&-\frac{\partial H_z(x,y,t)}{\partial x}=j_{y}(x,y,t)+\epsilon_0\frac{\partial E_y(x,y,t)}{\partial t} +\frac{\partial P_y(x,y,t)}{\partial t}.\nonumber\\\nonumber
\end{eqnarray}
The $j_x,j_y$ are components of electric current density, $\epsilon_0$, $\mu_0$ are vacuum permittivity and permeability. The polarization components $P_x,P_y$ are calculated wit a separate equation \cite{Alsunaidi}
\begin{equation}\label{polaryzacje}
\ddot{P}+\gamma\dot{P}+\omega^2_{0} P=\frac{\omega^2_{p}}{\epsilon_\infty} E
\end{equation}
with material parameters $\omega_p$, $\gamma$, $\epsilon_\infty$ characterizing the given optical medium. In the frequency domain, the above model yields susceptibility $\epsilon(\omega)$ in the form 
\begin{eqnarray}\label{drude}
\vec{P}(\omega)&=&\epsilon(\omega)\vec{E}(\omega),\nonumber\\
\epsilon(\omega)&=&\epsilon_\infty+\sum_{j=1}^{n}\frac{\omega_{pj}^2}{\omega_{0j}^2 - \omega^2 - i\gamma_j\omega}
\end{eqnarray} 
where a set of oscillators $j=1,2...$ is used, each with corresponding equation (\ref{polaryzacje}), in order to model the optical response of the medium. 

\section{Appendix B: Material models}
\begin{figure}[ht!]
\includegraphics[width=.95\linewidth]{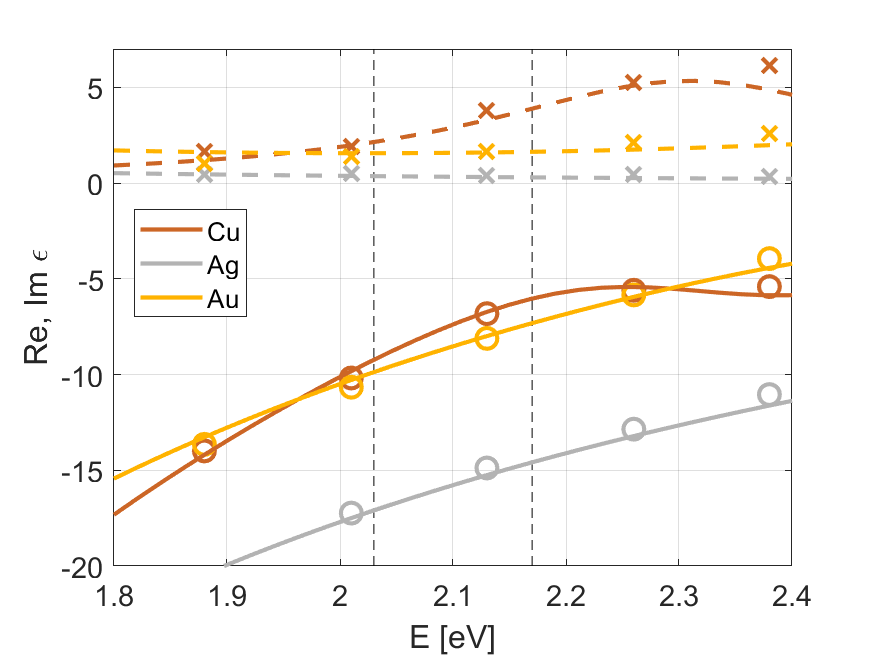}
\caption{Real and imaginary parts of permittivity from literature data \cite{Hollstein,Johnson} (dots) compared to numerical model (lines).}\label{fig:Cumodel}
\end{figure}

In this paper, we consider three metals commonly used in plasmonics: gold, silver and copper. A comparison of the real and imaginary part of susceptibility obtained from the model used in calculations with experimental data from \cite{Hollstein} and \cite{Johnson} is shown on Fig. \ref{fig:Cumodel}.

The region of interest is the area between two dashed lines indicating the wavelength corresponding to $1S$ exciton and band gap; the entire yellow exciton spectrum is located within these limits. As the wavelength approaches the band gap, for copper and gold Re $\epsilon \rightarrow -7.5$; this indicates that highly focused plasmons at Cu-Cu$_2$O interface can be excited by the light with this wavelength.

\end{document}